\documentclass[conference,compsoc]{llncs}

\usepackage{graphicx}
\usepackage{xcolor}

\usepackage{amsmath}
\usepackage{amssymb}
\usepackage{dirtytalk}
\usepackage{todonotes}
\usepackage{caption}
\usepackage{subcaption}
\usepackage{xspace}

\usepackage{tikz}
\newcommand*\circled[1]{\tikz[baseline=(char.base)]{
            \node[shape=circle,draw, fill=white,inner sep=1pt] (char) {#1};}}

\newcommand{\D}{I2D\xspace}
\newcommand{\Ds}{I2Ds\xspace}
\newcommand{\F}[1]{\mathcal{#1}}
\newcommand{\Fn}[1]{\mathcal{#1}}
\newcommand{\Fd}[1]{$\mathcal{#1}$}
\newcommand{\M}[1]{$\mathbb{#1}$}

\title{Information Inference Diagrams: Complementing Privacy and Security Analyses Beyond Data Flows}

\author{Sebastian Rehms\inst{1} \and Stefan Köpsell\inst{1,2,3} \and Verena Klös\inst{4} \and Florian Tschorsch\inst{1}}
\authorrunning{Rehms et al.}
\date{October 2023}
\institute{Technische Universität Dresden, Germany (\email{first.last@tu-dresden.de})
\and
	Centre for Tactile Internet with Human-in-the-Loop (CeTI), Dresden, Germany
\and
        Barkhausen Institute, Dresden, Germany (\email{stefan.koepsell@barkhauseninstitut.org})
\and
        Carl von Ossietzky Universität
Oldenburg, Oldenburg, Germany (\email{verena.kloes@uni-oldenburg.de})
 }
 
\begin{document}

\maketitle

\begin{abstract}
This work introduces Information Inference Diagrams (\Ds),
a modeling framework aiming to complement existing approaches
for privacy and security analysis of distributed systems.
It is intended to support established threat modeling processes.
Our approach is designed to be compatible with Data Flow Diagrams~(DFDs),
which form the basis of many established techniques and tools.
Unlike DFDs, \Ds represent information propagation,
going beyond mere data flows to enable more formal reasoning
in threat modeling while remaining practical.
They define inference and sharing (flow) relations on information items
to model how information moves through a system.
To this end, we provide formal definitions
for information items, entities, and flows.
By introducing classes as a type system, our formal rules are both generic
and allow conformance to existing vocabularies.
We demonstrate the applicability of \Ds through examples,
that showcase their versatility in system analysis.

\end{abstract}

\section{Introduction}
Security and privacy analyses often begin with a description of the target system.
These descriptions form the foundation for systematic analysis,
enabling tasks such as identifying risks associated with system elements
or verifying compliance with specific requirements.
In particular, Data Flow Diagrams (DFDs) have gained widespread adoption for practical threat modelling \cite{shostack_threat_2014}.
Techniques such as STRIDE~\cite{stride_2009} and LINDUN~\cite{deng_privacy_2011}
employ DFDs systematically and have been extended
for specialised scenarios and contexts~\cite{leicht_creating_2024,sion_solution-aware_2018}.
Furthermore, these methods have been integrated into tools
for threat modelling~\cite{noauthor_owasp_nodate,noauthor_iriusrisk_nodate,sion_sparta_2018}.

While data-flow-centric approaches are productive for assessing privacy and security,
they often fail to address deeper, fundamental aspects of security and privacy:
protecting data is often the means to an end;
safeguarding the information implicitly or explicitly derived from the data
and its surrounding context.
Information, in this context, refers to the interpretation of data,
which can include unintended insights drawn from metadata,
contextual factors, or background knowledge.
Please note that this is a qualitative notion of
information~\cite{oliver_privacy_2014}
and not quantitative like entropy.

\begin{figure*}[t]
\centering
\includegraphics[width=.8\textwidth]{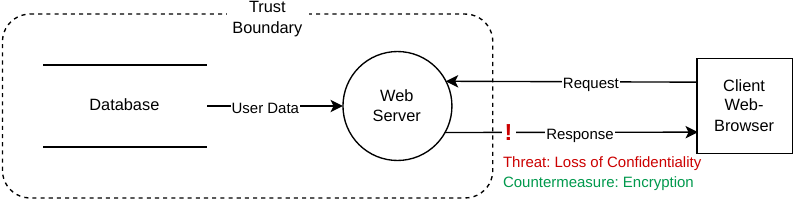}
\caption{Data Flow Diagram of a typical web application (request-response scenario).}
\label{fig:dfd}
\end{figure*}

DFDs, while effective for visualizing data flows,
are not designed to explicitly represent the information level of a system.
This limitation makes it difficult to analyse how information \textit{propagates}
through a system or to evaluate information access, especially
in terms of contextual or background knowledge.
Practical analyses using DFDs often address these gaps only implicitly
through annotations or ad hoc interpretations.
In Fig. \ref{fig:dfd}, we provide a simple example
illustrating some of the discussed problems.
It depicts a DFD of a distributed application
delivering data to a user via a web app,
with an annotated threat and countermeasure:
the response contains confidential user data at risk.
This annotation relies on additional, implicit knowledge.
It is implied that data is confidential, that data is included in the request
(despite no explicit link between the database and response),
transmitted over an untrusted channel, and mitigated by encryption.
It is also unclear, how adding encryption affects the system, e.g., if new data flows are necessary to manage keys and identities, if new risks like identifiability arise (as the mitigation is only indicated as an annotation). 
Such knowledge is not visible because DFDs are not designed for structured reasoning.
DFDs are popular partially because they abstract from these kinds of details.
However, given higher  demands on precision and granularity of a threat modelling process, it becomes necessary to facilitate such aspects within a data flow model.
It is also possible that only smaller critical parts of a system need to be analysed in depth.

In this paper, we address these challenges
by introducing \emph{Information Inference Diagrams (\Ds)},
a novel methodology for reasoning about privacy and security.
\Ds provide a structured representation of a system's information view,
enabling a clearer understanding of how information propagates
to facilitate the analysis of security and privacy properties.
Our approach is designed to complement existing methodologies,
such as DFD-based threat modelling,
by bridging the gap between data-level and information-level analysis.
\Ds are intended to fill the space between formal information flow analysis \cite{mantel_information_2011} and classical threat modelling.
They are intended to be used in DFD-based analysis to enable detailed inspection.

To this end, \Ds provide a framework for modelling
the existence of \textit{information items} within a system,
their propagation pathways, and the inferences that can be drawn from them.
This enables \Ds to capture and analyse
the impact of successful attacks or architectural changes on the system.
\Ds are specifically designed to allow direct translation from DFDs,
enabling integration with established system modelling practices.
They follow a modular approach,
allowing focused analysis on selected parts of a system while abstracting others.
An important property of \Ds is their support for incremental model construction,
where additional information can be iteratively incorporated by the user.
\Ds can be either constructed manually or automatically generated
from existing system descriptions.
However, the automatic generation of \Ds,
which depends heavily on the specific characteristics of the input model,
is beyond the scope of this paper.

The contributions of this paper can be summarised as follows.
We introduce Information Inference Diagrams (\Ds) by providing definitions for all elements and by explaining certain concepts making them versatile for complementing existing DFD-based analysis endeavours.
Furthermore, we demonstrate the utility of \Ds through illustrative examples,
showcasing their effectiveness in identifying and mitigating
security and privacy risks in various scenarios.

The remainder is structured as follows.
In Section~\ref{sec:def}, we define \Ds and their foundational concepts.
In Section~\ref{sec:application}, we demonstrate their application through examples.
In Section~\ref{sec:discussion}, we discuss key challenges and design decisions.
In Section~\ref{sec:relw}, we review related work and position \Ds in context.
Lastly, in Section~\ref{sec:conclusion},
we conclude and propose directions for future research.

\section{Information Inference Diagrams}
\label{sec:def}

In this section, we introduce the foundations of \Ds
and their construction principles.
\Ds are intended to complement DFD-based reasoning by switching to information level relations.
Therefore they can build on existing DFDs.

\begin{figure*}[b!]
\centering
\includegraphics[width=.8\textwidth]{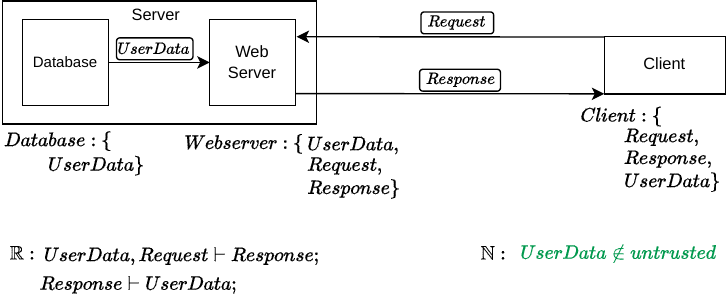}
\caption{A representation of the example system from the Introduction.}
\label{fig:idfd}
\end{figure*}

Fig. \ref{fig:idfd} shows an \D for the scenario depicted in Fig. \ref{fig:dfd}.
Independent of their type, all nodes are represented as \textit{entities} sharing information through \textit{information flows}.
The trust boundary is also an entity containing sub-entities.
Each simple entity has associated a set of \textit{information items}.
\textit{Inferences rules} ($\mathbb{R}$) describe the inference relation of these items.
Lastly a \textit{normative} requirement rule ($\mathbb{N}$) describes pricacy or security requirements.
In the following we define all these elements and provide syntax rules.
We will refer back to this example in Sec. \ref{sec:application}, when all foundations have been layed out.

\subsection{Elements}
At its core, an \D represents entities and the information that is available to them.
Information can become available to an entity by learning it from another entity (represented through flows) or by inference.

\Ds contain descriptive and optional normative declarations. Normative declarations (\M{N}) express requirements which should hold.
The descriptive part is composed of three aspects:
First, the information available to different entities.
Second, the flow of information between entities.
Third, inference rules (designated by~\M{R}) which can be automatically evaluated during the reasoning process.

\subsubsection{Entities and Information Items.}

The core of an \D are entities. Each entity $e$ possesses an individual information item set.
$\Fn{I}(e)$ returns this set of all information items of an entity $e$.
An information item is an object representing some fact.
For example: given some message different information items can be defined, referring to aspects of the message itself, inferred properties of that message or context of the message.
One may define information items for the content of a message or the length of the message.
Regarding the security of the message, one can define an item referring to the inferred state of the message's integrity.
Similarly, one may define an item representing the time the message was received.
An information item can represent arbitrary knowledge for an entity and we make no restrictions on that, which enables to express a variety of security and privacy properties.

$\Fn{I}(e)$ is a fuzzy set, allowing to express uncertainty on information items.
Information items can be classified meaning that they can be annotated with additional class symbols assigning them to classes (e.g., an item may be classified as \say{personal data} or \say{high clearance level}).
We call the set of all information item classes $\mathfrak{C}^I$.
We define an information item as a tuple:
\begin{equation*}
i = (n, p, C),
\end{equation*}
 where $n$ is a unique name (identifier), $p$ is the grade of membership, and $C \subseteq \mathfrak{C}^I$ is a set of item classes the item is assigned to.
 As a shorthand notation, one can write $n^{c_0,...,c_n}_p$ where $c_0,...,c_n$ are the class names in $C$.
 If $p = 1$, it can be omitted.
Classes work like annotations for elements, allowing to define group-based rules, which we introduce later.
The function~$C(i)$ designates the classes of some item $i$.

\Fd{E} is the set of all entities. An entity can be simple or complex, where
\Fd{E}$^s$ is the set of all simple entities and \Fd{E}$^c$ the set of all complex entities.
A simple entity is a tuple:
\begin{equation*}
e=(n, I, C),
\end{equation*}
where $n$ is a name, $I$ is a set  of information items, and $C \subseteq \mathfrak{C}^E$ is a set of entity classes ($\mathfrak{C}^E$ refers to the set of all possible classes for entities).
A complex entity consists of a name, classes, and child entities.
Hence an entity $E$ in general is constructed recursively:
\begin{equation*}
E = e \mid (n,C,{E}),
\end{equation*}
where $e$ is a simple entity and $(n,C,{E})$ is a tuple representing a complex entity.
$n$ is a name, ${E}$ is a set of sub-entities and $C$ a set of entity classes.
$C(e)$ designates the class set of an entity $e$.
The recursive construction generates a tree like structure in which simple entities are the leaves.

In \Ds, the notion of an entity is very general.
It may represent physical components (e.g., a router), virtual components (e.g., an API endpoint), humans, or physical and virtual channels.

\subsubsection{Flows.}

\Fd{F} represents the set of all flows.
A flow can be simple or complex.
A simple flow consists of an origin, a target, associated information, a name, and classes:
\begin{equation*}
f = (n,o,t,I,C)
\end{equation*}

The construction of a flow is similar to entities:
\begin{equation*}
F= f \mid (n,o,t,C, {F}),
\end{equation*}
where
$n$ designates the name of a flow.
$o,t \in \mathcal{E}^s$ and $o \neq t$  designate the origin and target of a flow which are always simple entities.
$C \subseteq \mathfrak{C}^F$ designates classes for flows, where $\mathfrak{C}^F$ is the set of all flow classes.
${F}$ is a list of sub-flows.
$I$ is a set of information items associated with a flow.

Flow-associated items express that information available at the origin is shared with the target.
Let $i$ be any information item, $a,b$ simple entities and $f$ a simple flow from $a$ to $b$ with associated information items $\F{I}(f)$, then:
\begin{equation*}
i \in\F{I}(a),\,i \in \F{I}(f) \rightarrow i \in \F{I}(b).
\end{equation*}

Note, that a flow does not necessarily mean, that information is sent from one entity to another; a flow rather represents that \textit{only if} some information is available at one entity this information is also available to the other.
We allow the wildcard symbol~$*$ as information item on a flow to express that one entity shares all its information with another entity.

Complex flows allow to specify the structure of information exchange and can therefore be used to model protocol-like interactions.
Complex flows contain a list of possibly complex children flows.
As every flow has an origin and a target entity, complex flows logically also contain associated entities, denoted by $\F{E}(F)$, where $F$ is a complex flow.
These entities can be \say{external} to the original flow, i.e., they are not required to be the origin or target.
Helper entities therefore can be introduced to describe complex flows.
In fact, complex flows can be seen like nested diagrams.

\subsubsection{Rules.}

New information items have to be added to the information sets of entities if inference rules in~\M{R} apply.
In the simplest form, these rules follow the syntax
\begin{equation*}
a_0,a_1,... \overset{C}{\vdash}_p b_0,b_1,...
\end{equation*}
Here, $p$ is a probability $\in [0,1]$.
$a_0,a_1,...$ and $b_0,b_1,...$ are information items.
$C \subseteq \mathfrak{C}^E$ are classes of entities, to which the rule applies; it can be omitted implying that it applies to all entities.

If all information items~$a_0,a_1,...$ are present in an entity set the rule applies, triggering an addition of the items~$b_0,b_1,...$ with degree $p$ to the respective entity set ($p$ may be omitted implying $p=1$).

Class-based rules can be written according to this syntax:
\begin{equation*}
        c_0(x),c_1(y),... \overset{C}{\vdash} b_0,b_1,... 
           \qquad\text{or}\qquad 
    c_0(x),c_1(y),... \overset{C}{\vdash} x^{d_0},y^{d_1},...  
\end{equation*}
where $c_0, c_1,..., d_0, d_1,... \in \mathfrak{C}^I $ are item classes and $x,y$ are variables (must not be bound by an item name).
The first rule schema conditionally adds items.
The second rule schema serves the purpose to conditionally rewrite classified items, e.g. the rule $c0(x),c1(y) \vdash x^{d_0},y^{d_1}$ would apply to the item set $\{a^{c_0},b^{c_1}\}$ and generate the new items $a^{d_0}$ and $b^{d_1}$.

It is important to prevent infinite recursion.
This can be done by implicitly marking the outcome by the generating rules and prohibit re-application.

Classes in rules can be negated (effectively yielding the complement set): $c(x),\neg d(x) \vdash i$ applies if $c \in C(x) \wedge d \notin C(x)$.

For class rules, one can use the variable~$x$ in the outcome, which can be useful, e.g., to \say{rewrite classes}.
For instance, if $c$ is a class, then the rule $c(x) \vdash x^{c,d}$ would add a new item with classes $c,d$ to $x$.

A wildcard class $*()$ can make a single-class-rule to be applied to multi-classed items.
For example, $c(x),*(x) \vdash j$ applies to an information item $i^{c,d}$, although $d$ is not explicitly mentioned in the rule.
Using a class variable, indicated by the wildcard $*$, allows to propagate the classes to the outcome: $c(x),*v(x) \vdash j^v$.

An arrow can be used to express replacement of items on all outgoing flows.
If $i$ and $j$ are information items, $c$ is class and $x$ a variable, then the two rules
\begin{equation*}
    i \overset{m}{\rightarrow} j
       \qquad\text{or}\qquad 
    c(x) \overset{m}{\rightarrow} j
\end{equation*}
express that all outgoing flows containing $i$ or $c$-classified items instead contain $j$ but only if $j$ is available for the entity the rule is applied for (entities classified $m$).
Such rules can be used, to automate the expression of transformations of items at intermediate stations (like pre-processing etc).
In Fig.\ref{fig:c3} an example is given how such a rule in combination with other inference rules can be used to represent man-in-the-middle-like attacks.

Entity classification rules allow to automatically add items to information sets of entities, add new flows or child-entities based on classification.
If $c$ is an entity class symbol, $i$ an information item, $E$ is an entity and $f$ is a simple flow , then one can use:
\begin{equation*}
C(c) \vdash i
   \qquad\text{or}\qquad 
C(c) \vdash e
   \qquad\text{or}\qquad 
C(c) \vdash f
\end{equation*}
The first means that all entities of class $c$ possess $i$.
The second means, that if an entity is of class $c$, then it automatically has a child $E$, transforming the parent entity to complex one if it is not already one (note that we allow $E$ to be complex itself).
The third means that a $c$-classified entity is part of a flow.
For flows we force that the respective entity is either the origin or target of the flow.
If the other entity does not exist, it is created.
Note that the associated information items are not required to be in the entities' sets (cf. the definition of flows above).
Similar, rules allow the automatic addition of nested entities and flows.

One aspect to heed is, that automatic changes triggered by class rules may add items to complex entities.
This is problematic because complex entities can be classified but do not possess their own information sets (only the leaf entities).
Therefore in such cases the triggered rules need to be resolved to add the according elements to leaf elements.\footnote{The alternative would be to differentiate between specific classes for simple and complex entities. This, however, would also make a resolving step necessary, as the class of a simple entity needs to be transformed into the class of a complex entity when adding sub-entities.
Our approach fits together with the concept of views introduced further down.}

\subsubsection{Normative Declarations.}
All aspects presented so far are purely descriptive,
as they focus solely on enabling the representation of a system.
By introducing normative requirements,
it becomes possible to evaluate whether any issues exist
within the current system model.
This step is entirely optional.

Requirements follow the patterns:
\begin{equation*}
a,b,... \in E
   \qquad\text{or}\qquad 
a,b,...  \notin  E,
\end{equation*}
where $a,b,...$ are information items,
and $E$ is a simple or complex entity.
One can use a wildcard $*$ to represent every entity
and add exceptions using $\setminus F,G$ for entities $F,G$.
Requirements state that an entity shall or shall not have access to specific information items.
If $E$ is a complex entity, this implies that the respective information item is/is not present at any/at least one sub-entity (cf. the definition of views).
Instead of $E$, one can also use an entity class, effectively making it possible to declare policies for sets of entities.

\subsection{Transformations}
Recall that models are inherently incomplete,
which is why we focus on the construction process of \Ds.
In this section, we describe how an \D,
which can be generated from existing system models or DFDs,
can be further refined and specified through transformations.

Transformations refer to any change of the model during the modelling process.
Rules explained above can be understood as automatic transformations to the model.
But during them modelling process, the user introduces also changes according to the possibilities of the modelling method.
This can be understood as manual transformations, adding additional knowledge to the model.

Simple transformations include the addition of information items,
the introduction of simple entities and flows,
the classification of existing elements,
and the creation of rules.

More elaborate transformations involve refining entities
by adding sub-entities or refining flows,
both of which result in the creation of complex elements.
When a simple entity is transformed into a complex entity
through the addition of nested sub-entities,
incoming flows directed at the parent entity
must be rewritten to target specific child entities.
This adjustment aligns with the definition of flows,
which do not permit complex entities as origin or target.

For the creation of a complex flow,
we specifically introduce the concept of a \textit{bisection}.
This manual transformation answers the question: \textit{how is the information exchange mediated?}
A bisection introduces a mediating entity in between two information sharing entities.
This new entity represents the logical path enabling the sharing of information.
Depending on the scenario, the mediating entity can be very general (e.g., an ISP) or very specific (e.g. a physical channel).
A bisection implies that the information set of the mediating entity contains all information items belonging to the flows---otherwise it would not be mediating the information.

The result of a bisection is a complex flow which can be specified further, e.g., by adding other entities or flows.
For example, the ISP entity might be composed of multiple routers;
the physical channel might imply that other entities can eavesdrop on it and therefore a new flow can be added.
A bisection is used as example in Fig. \ref{fig:c1}.

Each transformation possibly triggers a cascade of changes and therefore an iterated re-evaluation of rules is necessary.

\subsection{Views}
\begin{figure*}[tbp]
\centering
\includegraphics[width=1\textwidth]{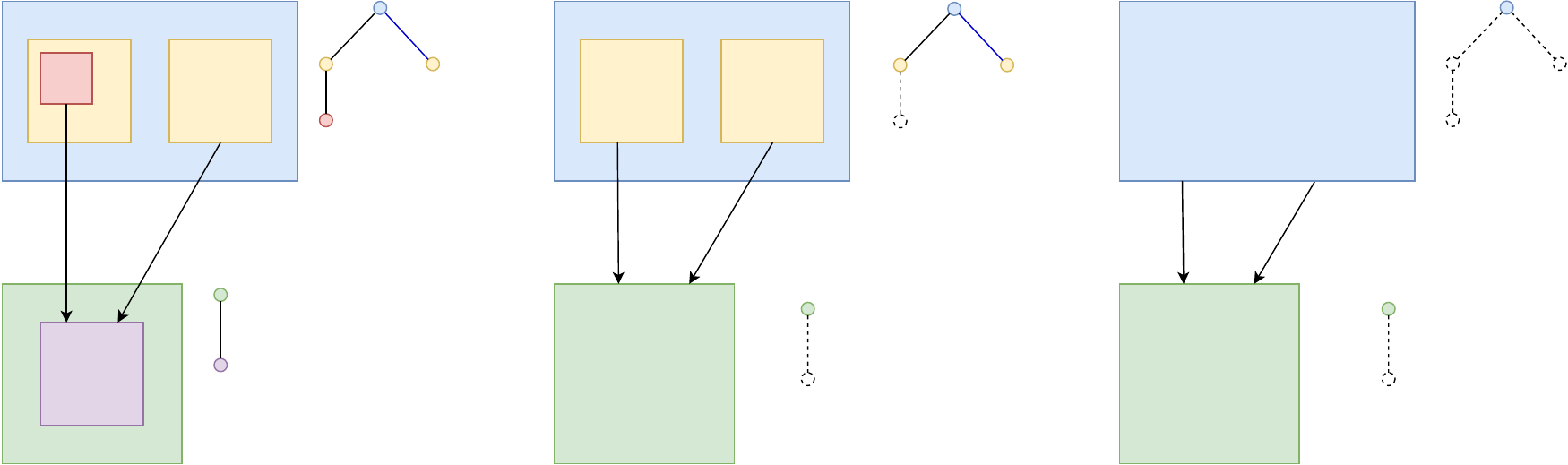}
\caption{Views on two entities (without information items).
The tree representation next to each entity shows the abstracted sub-tree.
Colours indicate different possible views, which may vary between entities.}
\label{fig:zooms}
\end{figure*}

A view of a complex entity represents its abstraction
at a specific depth of the tree,
reducing it to a simple entity.
This is achieved by recursively \say{folding} all information items
from its child entities into the into the parent entity's information set.
In essence, a view abstracts all nested entities
by creating the union of the information sets of all leaf nodes in the hierarchy
(using fuzzy set standard union, selecting the highest probability).
A complex entity thus aggregates the information of its children.
Flows directed to the child entities are redirected 
to the newly instantiated simple entity.
Figure~\ref{fig:zooms} illustrates this process:
Nested boxes represent entities and the colors represent possible views.
Tree representations next to each complex entity show abstracted components.
Left to right depicts three levels of abstraction,
omitting specific information items.

Views serve two main purposes:
First, they aim for complexity reduction by abstraction.
Second, they can be interpreted as inverted specification:
Abstracting a view hides the specification done for the respective element.
Turning this around allows us to state: simple elements can be understood as not yet specified, emphasizing incompleteness of models.

Views of complex flows are not straightforward,
as such flows may involve multiple entities and additional sub-flows.
A complex flow can be represented as a simple flow by removing all sub-elements
from the diagram's representation.
Alternatively, each complex flow can be \say{deconstructed,}
as a complex flow can be interpreted as a named or classified section of the \D.

\subsection{Schemata}
All elements can be predefined as a schema
before starting the construction of an \D.
Such schemata can support the analysis
by ensuring adherence to a specific vocabulary
or by forcing the analysis to address particular requirements.

For instance, a schema may define protocol symbols
and provide corresponding inference rules for those protocols.
The class $TCP$ could be predefined with inference rules
such as $TCP \vdash fingerprinted\_operating\_system$,
or the class $TLS \vdash authenticated$.
Another example could involve the classification
of information items and entities
according to European GDPR properties (similar to~\cite{grunewald_tilt_2021}).
\Ds in this sense can be used as kind of a meta-language.

\section{Application Examples}
\label{sec:application}
In this section, we provide application examples,
illustrating how \Ds can be used for analysing systems.
The main purpose is to highlight the use of different properties and why they have been defined the way they are.

\subsection{Example 1: Request/Response Scenario}
Here, we refer back to the contrasting example from Fig. \ref{fig:dfd} and Fig. \ref{fig:idfd}, showing a possible representation of the DFD-based system description as \D.
The original trust boundary is represented by a complex entity (e.g., a server running two services).
Note that this is an interpretation of the underspecified situation in the original DFD.
(We will the topic of trust boundaries later in Sec.~\ref{sec:discussion}).
Each entity owns its own information set.
The database shares the item \emph{UserData} with the web server through a flow.
As the item \emph{UserData} is an element of the information set of the database, it is also an element of the information set of the web server.

Another flow represents the client request.
In the original DFD, the \say{Request} expresses that some data flows across the trust boundary.
Here, the item \emph{Request} is an abstract item representing the data contained in the request that is initiated by the client.
The \emph{Request}-item is an element of the information set of the web server because of two facts:
the flow, stating, that the information is shared if it is present at the origin and the fact that it is indeed present at the client (the client possesses the information of the $Request$ by definition).

The same holds to the response flow:
in the \D, the server shares the \emph{Response} item with the client only,
if it is part of its information set.
A rule allows the server to \textit{infer} a response from both the \emph{Request} item and the \emph{UserData} item from the database.
\Ds prompt to make such connections between information items explicit.
The flow applies and the \emph{UserData} is available at the client.

The annotated threat in the DFD is represented through a normative rule in \M{N} (in green).
The rule states that \emph{UserData} must not be an element of any entity classified as \emph{untrusted}.
This is the case so the rule is not violated.
This rule is the equivalent of a \textit{confidentiality requirement} on a concrete item.
Any confidentiality requirement can be stated like that as it precisely encodes confidentiality: only authorised parties get access to some information.
We will address the relation of protection goals and threats in \Ds later in Sec.~\ref{sec:discussion}.

\subsection{Example 2: Complex Analysis}
\begin{figure}[tbp]
\begin{enumerate}
    \renewcommand{\labelenumi}{(\alph{enumi})}
    \item Content provided from server to client:\\[1em]
     \begin{subfigure}[b]{\textwidth}
        \includegraphics[width=0.9\textwidth]{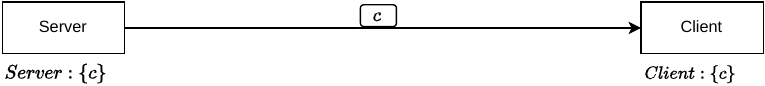}
        \phantomcaption\label{fig:c0}
    \end{subfigure}

    \vspace{1em}
    
    \item Bisected flow mediated by ISP:\\[1em]
    \begin{subfigure}[b]{\textwidth}
      \includegraphics[width=0.9\textwidth]{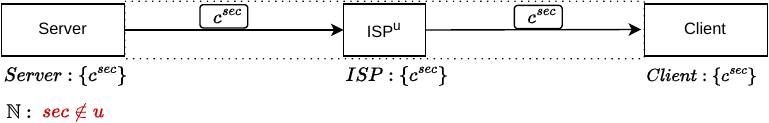}
      \phantomcaption\label{fig:c1}
    \end{subfigure}

    \vspace{1em}

    \item Simply encryption added to prevent content leak:\\[1em]
    \begin{subfigure}[b]{\textwidth}
        \includegraphics[width=0.9\textwidth]{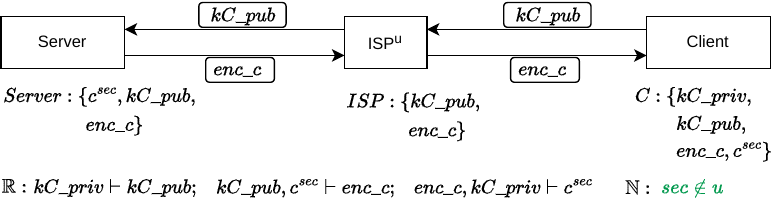}
        \phantomcaption\label{fig:c2}
    \end{subfigure}

    \vspace{1em}

    \item Representation of a Man-in-the-Middle attack using class-based rules:\\[1em]
    \begin{subfigure}[b]{\textwidth}
        \includegraphics[width=0.9\textwidth]{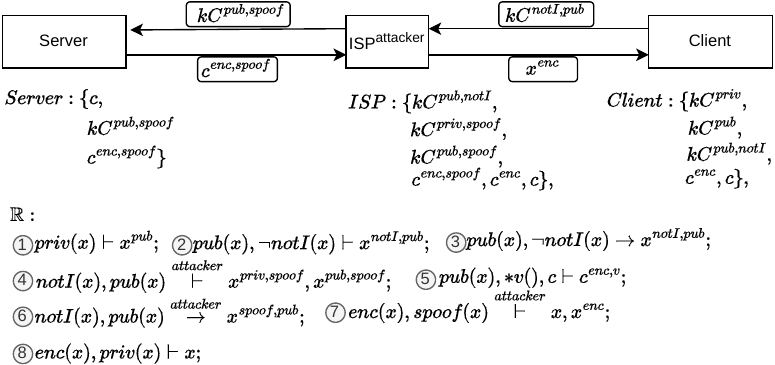}
        \phantomcaption\label{fig:c3}
    \end{subfigure}

\end{enumerate}
\caption{Development of a Server-Client relation.}
\label{fig:c}
\end{figure}

The next example illustrates how our method can be used in different ways to analyse a system.
The example is a generic version of the first one, representing a typical client-server relation (Fig. \ref{fig:c}, where the entity \say{Server} is providing content for a generic \say{Client} (see Fig.~\ref{fig:c0})).
One can think of the server as an abstracted view on the complex entity from the first example as we focus on the relation to the client.
The server holds the information item $c$ representing some content.
Through the flow (for simplicity we omit names for flows in this example) the information item $c$ is available for the client too.

Fig. \ref{fig:c1} presents a flow bisection in which the ISP is introduced as \textit{mediating entity}.
The complex parent flow is indicated by the dotted line.
Mediating entities need to know the mediated information (therefore $c$ is in the newly introduced information set of the ISP).
Additionally, two class annotations are introduced: $sec$ for the content $c$ expressing that it is secret and $u$ for the untrusted ISP-entity.
Therefore, the normative requirement states that $sec$-classified items should not be available to $u$-classified entities.

Fig. \ref{fig:c2} represents a transformation introducing a simple encryption to prevent the rule violation.
Multiple steps lead to the shown representation.
First, an item $kC\_priv$ is added to the information set of the client, representing a private key.
The first inference rule in \M{R} applies to the client and adds a public key $kC\_pub$ to the information set of the client.
A new flow is introduced, transferring a public key of the client ($kC\_pub$).

ISP learns $kC\_pub$ and another flow adds it to the information set of the server.
The second rule applies, deriving the encrypted content, which is propagated to the client.
The last rule allows the inference of the content.
The normative requirement is not violated.

Fig. \ref{fig:c3} analyses how a man-in-the-middle attack can be represented.
For didactic purposes, we switch here to class-based rules.
$pub$ and $priv$ mean that the item is of class public and private key material, $enc$ that something is encrypted and $notI$ means that there is no integrity protection on the item.
Other possibilities for representing integrity are possible, e.g. by introducing an integrity information atom which can be inferred from the private key.
A \emph{spoof}-class expresses that something has been altered by an \emph{attacker}-classified entity.
$kC$-items refer to client key material.

Rule \circled{1} generates the public key from the private one.
\circled{2} generates a public key which is not integrity protected (the rule applies to $kC^{pub}$).
\circled{3} uses a rewrite arrow, ensuring that the not-integrity-protected key is sent.
Rule \circled{4} generates a spoofed public and private key -- note that rule \circled{1} would not apply, as it does not explicitly state $notI$ as target class.
\circled{5} exemplifies the wildcard syntax: an encrypted variant of any information atom with name $c$ can be inferred, independent of its classes; the only additional requirement is that some other $pub$-classified item is present.
The rule applies at the ISP as well as the server, in both cases generating $c^{enc,spoof}$ and only for the ISP $c^{enc}$.
\circled{6} is a rewrite rule, ensuring that an attacker entity (here the ISP) is indeed trying to replace unprotected public keys with a spoofed version.
\circled{7} enables the attacker to generate the client-decryptable version of the content.
Lastly, \circled{8} enables the client to regularly decrypt by using its private key.

The whole attack could be represented in many other different and simpler forms.
We mainly want to show how class-based reasoning may operate.

\section{Discussion}
\label{sec:discussion}
In this section, we  discuss several design decisions as well as advantages and limitations of our approach.

\subsubsection{Translating DFDs to \Ds.}
DFDs contain several specific aspects which are not directly encoded in \Ds.
The different elements in a DFD can be interpreted as encoding different states of data:
for instance a database implies data at rest while a process implies data in use.
In our model, classes of entities and classes of items can be used to represent such aspects.
Additionally, data in transit can also be modelled through the rewrite operator.

Trust boundaries are a specific aspect in DFDs of which several interpretation are possible.
From the point of a security analysis the significance lies in a hint for the analyst to be especially careful with data flows crossing a trust boundary as they originate from a lower trust environment, e.g. making an I/O validation appropriate to prevent injection attacks.
As seen in Fig. \ref{fig:idfd},  a trust boundary can simply be modelled as a complex entity;
again, if necessary, one may use specific classes for such special collections of entities.
In \Ds, one can also add according normative requirements to ensure a check for specific validation inferences on items from untrusted entities,
e.g. a normative rule may be added that requires an I/O validation or escape item to be present.

\subsubsection{Reasoning during the Security Analysis Process.}
As stated above, we want to provide a method which can be used, given a DFD, to enable structured reasoning.
This can be especially viable, when mitigations are to be added to a system, which often implies actually adding additional information to a system --- for example when adding cryptography, keys need to be managed.
In particular from a privacy perspective, it may be appropriate to reason exactly what effects the additional information can have; e.g. if there are new privacy relevant inference relations possible.

Additionally, our modelling method can be used to express specific circumstances of a system, e.g. in Fig.~\ref{fig:c3} a MitM-attack.
This is why we state that our tool allows structured reasoning in the greater context of security analysis:
the elements of \Ds allows to show exact information flow and inference conditions, which lead to the realization of a threat, and hence to analyse the effects of counter measures.
The main features we consider relevant in this context are explicit information binding to entities, including the information propagation and the inference relation to understand where which information actually originated from.

As models are inherently incomplete, their core benefit lies in abstracting unessential parts for a specific interest and accentuating important ones.
The challenge therefore lies in finding the \textit{relevant} aspects.
And while a system model can theoretically be taken as a fixed structure, it is always created in a process -- and can hence also be understood as structured reasoning about system model parts (a change implies the statement that the change is actually adding something relevant to the model).
Given this perspective, we included specific ways to transform \Ds to support the search for aspects especially relevant to privacy and security:
the nesting of entities and flows and the bisection operation as special complex flow constructor.
These actions are intended to hint for privacy and security specific questions like:
How is it possible that an information item is available at another entity?
What are the components of a data-processing entity and do they imply additional information and inference possibilities?
Are there more relevant subcomponents implying new flows?

\subsubsection{Relation to Threat Modelling.}

One important facet is the relation of threats and \Ds.
A requirement is not necessarily connected to a threat (depending on the notion) --- but every threat implies a requirement.
Our modelling approach intends to clearly differentiate between normative and descriptive aspects;
this forces the user to make the requirements behind threats explicit.
In \Ds, every requirement has to be represented through \textit{element relations} on items and entities (cf. the definition of normative declarations in Sec \ref{sec:def}).
Most protection goals can be represented like this if one understands an information item as a \textit{capability} of inference or the absence of possible inferences.
For example, integrity can be understood as the possibility to infer the state of integrity of something (e.g. with a rule $privKey, message \vdash i\_m$).

Threat modelling approaches address the incompleteness problem by providing threat categories which could possibly apply.
They are often very generic and still require a lot of additional knowledge.
To solve the issue in practice, LINDDUN, for example, provides ~\cite{deng_privacy_2011,sion_robust_2025} threat trees for privacy threat categories.
Such threat trees can be understood as extending the catalogue of threats by specific sub-catalogues of possible realizations.
This helps during the modelling process to look for specific kinds of problems, supporting (but not ensuring) completeness.
Such catalogues can also be used in our modelling method, sometimes this might actually be recommendable.
\Ds are intended not to replace established approaches in threat modelling but to complement the process.
In \Ds the user is required to be more specific about the exact condition, how a threat may manifest itself in relation to a requirement.

For instance, in LINDDUN, the threat of detection can be realized through the sub-threat of observed communications.
In our model it has to be clear which entity might observe the traffic (which may be introduced through a bisection).
One can also add a possible information item \say{communication patterns}, an appropriate inference rule and a requirement prohibiting this item at the mediating entity.

\subsubsection{Limitations.}
We want to shortly address some limitations of our modelling approach.
All requirements have to be expressed in a specific way, which can be challenging in some situations.
For example, the problem of unawareness is not straight forward.
One solution may be to include it as a lack of integrity in the mental model of the user.
Availability is another problematic aspect, as it requires some kind of time-dependency on information items, making it necessary to define all information items with logical time indices (one variable for each logical time step).

This shows that \Ds are not a silver bullet but help analysing specific problems.
Every solution needs to strike a balance between generalization (at which DFDs are very good) and specificity, making them applicable for more general or specific situations (the latter is the case of \Ds).

One notable design decision is that there is no deletion of information items, making it hard to analyse a system in the sense of a data life cycle (as is required e.g. in~\cite{rost2024standard}.
The rationale here is twofold.
First, we want to avoid a system process model or state machine to simplify modelling.
Second, our approach intends to address hard security and privacy properties in which the loss of control over information is irreversible.
As stated, our approach is not intended to be a universal tool for privacy analysis but a complementing one.
Still, it is possible to express data deletion, if necessary, through classifying inference rules, which is, however not intuitive.

\section{Related Work}
\label{sec:relw}
Several works propose automation-focused approaches
to threat modelling~\cite{sion_automated_2021,van_landuyt_threat_2021},
including knowledge base requirements~\cite{wuyts_knowledge_2019}. 
These works focus on the general threat elicitation.
Especially DFDs have been the subject of several works,
addressing drawbacks like formal foundations~\cite{larsen_formal_1994},
lacking specific representation~\cite{sion_interaction-based_2018},
and automation~\cite{berger_automatically_2016}.
Other works propose DFD extensions
for special use cases~\cite{leicht_creating_2024,sion_solution-aware_2018}. 
Especially refinement and transformation similar to our ideas
have been proposed~\cite{alshareef_refining_2021}.
The research evolution of DFDs suggests a shift to a more information-centric focus,
as we propose here.

Other works focus on reasoning about security using security model languages,
e.g., \cite{holm_p2_2015,lodderstedt_secureuml_2002}).
However, these approaches lack flexibility,
as they require strict adherence to predefined vocabularies.
In contrast, our approach allows freely specifiable schemata and can in this sense be understood as a kind of meta-language.

A similar idea to ours has been proposed by~\cite{bavendiek_privacy-preserving_2018}, which models data propagation in a system
incorporating entity-bound information and probabilities.
It also differentiates between descriptive and normative declarations.
A key difference is that their approach uses entity-specific inference,
requiring propagation, whereas our inference rules are, at least without further classes, global by default.
This is an intentional design decision to underline a security insight: no security by obscurity -- which means that any entity can infer if it has the respective information.
Additionally, their work uses states to allow an order of events,
which we do not do for conformance and simplicity purposes.

\section{Conclusion}
\label{sec:conclusion}
In this work, we introduced the theoretical foundations of Information Inference Diagrams (\Ds). This framework enables modelling systems from an information-level perspective,
focusing on security and privacy analysis.
Unlike many established approaches centering on data,
\Ds provide a complementary tool for practical threat analysis.
By abstracting to the information level,
our approach enables to formally express security requirements
as part of the model through \textit{inference}.
We provided formal definitions and demonstrated the application of \Ds
through illustrative examples.

For future work, we plan to implement an automatic reasoning engine,
develop subject-specific schemata,
and integrate \Ds into existing threat modelling tools,
like OWASP Threat Dragon.

\section*{Acknowledgment}
Parts of this work has been supported by the following projects and funding bodies:
German Research Foundation (DFG) as part of Germany's Excellence Strategy (EXC~2050/1, Project~390696704, CeTI);
TU Dresden's Disruptions and Societal Change research framework (TUDiSC);
Federal Ministry of Education and Research of Germany (BMBF, Project~16KISA038, GANGES).

\bibliographystyle{splncs04}
\bibliography{SemiFormalSecAnalysis}

\end{document}